\documentclass[aps,preprint,nofootinbib,groupedaddress,showpacs,dvipdfmx]{revtex4}

\usepackage{graphics}
\usepackage{graphicx}
\usepackage{amssymb}
\usepackage{amsmath}
\usepackage{amsfonts}

\begin{document}

\title{Computational Method for the Retarded Potential in the Real-time Simulation of Quantum Electrodynamics}

\author{Masahiro Fukuda}
\author{Kento Naito}
\author{Kazuhide Ichikawa}
\author{Akitomo Tachibana} 
\email{akitomo@scl.kyoto-u.ac.jp}
\affiliation{Department of Micro Engineering, Kyoto University, Kyoto 615-8540, Japan}

\date{\today}

\begin{abstract}
We discuss the method to compute the integrals which appear in the retarded potential term for a real-time simulation based on QED (Quantum Electrodynamics). 
We show that the oscillatory integrals over the infinite interval involved in them can be efficiently performed by the method developed by Ooura and Mori based on the double exponential (DE) formula.
\end{abstract}

\pacs{11.15.Tk, 12.20.-m, 03.70.+k}



\maketitle

%
\section{Introduction}  \label{sec:introduction}
%

Quantum electrodynamics (QED) is a quantum theory of fields which describes the interaction between photons and charged particles, such as electrons and positrons.
The quantum field theory is conventionally solved using the covariant perturbation theory and this has succeeded in explaining many physical phenomena with very high accuracy. 
However, the perturbation theory is not suitable to follow the real-time evolution of the system.
It can only compute such quantity as a cross section, which just measures difference between the infinite past and infinite future, not the time evolution step by step.

This is not so satisfactory because recent experiments can probe shorter and shorter time scale about the ultrafast electronic dynamics in matter \cite{Krausz2009}. 
This includes photophysical and photochemical processes, and the real-time observation of such processes as photoemission from atoms, molecules, and surfaces is now possible at the attosecond order \cite{Pazourek2015}. 
Note that phenomena in which particle number changes, for instance photoemission, cannot be treated in a strict sense by quantum mechanics of point particles, and 
we need to use quantum field theory for their rigorous treatment. 
This is because, while particle number is conserved in quantum mechanics of point particles, 
quantum field theory provides a framework in which it is allowed to change.
Since photoemission involves photons, the quantum field theory we need is QED.
Therefore, in order to compare theoretical prediction with these time-resolved experiments and future more precise experiments, we need a real-time simulation based on QED, and, for that purpose, a non-perturbative method has to be developed.

Some ingredients required for such a method have been discussed in Ref.~\cite{Tachibana2015} by one of the authors.
There, it has been argued that, for the quantum field theoretic real-time simulation, it is not sufficient only to compute the time evolution of a wavefunction as is done in the quantum mechanics of point particles, but we also need to compute the time evolution of field operators (Heisenberg operators defined at each point in the spacetime) as well. 
These two types of time evolution are called the ``dual Cauchy problem" of quantum field theory in Ref.~\cite{Tachibana2015},
and we have to combine them to obtain time evolution of the expectation value of a quantum field operator corresponding to some physical quantity. 
Note that the Heisenberg ket vector is expressed by a linear combination of basis ket vectors whose coefficients are wavefunctions.
The basis ket vectors in turn are constructed by operating appropriate field operators on the vacuum ket vector. The time evolutions of the wavefunctions and field operators cancel each other to make the Heisenberg ket time-independent. 
This paper concerns one of the issues regarding the time evolution of the field operators which appear in QED.

In the literature, a method for solving quantum field theory in Heisenberg picture has been developed by Abe and Nakanishi \cite{Nakanishi2004}, with the special interest in the application for quantum gravity, in which the perturbation theory fails. 
Although the solution for QED is discussed in Ref.~\cite{Abe1992}, their method is quite formal and not convenient for our numerical approach. 
In particular, the solution in Ref.~\cite{Abe1992} is given as the expansion in powers of the electromagnetic coupling constant (electron charge magnitude), which seems to be not truly non-perturbative.
We wish to pursue a way to directly solve the quantum field equations of motion of QED by numerical means.

As for our approach, while theoretical developments are found in Refs.~\cite{Tachibana2001,Tachibana2003,Tachibana2005,Tachibana2010,Tachibana2013}, we have been developing a prototype code for a real-time simulation based on QED in Refs.~\cite{Ichikawa2013,Ichikawa2014,QEDynamics}. 
One of the important points we have dropped in Refs.~\cite{Ichikawa2013,Ichikawa2014} is the retarded potential term for the photon field. 
 In other words, we have regarded the photon field as a free radiation field and ignored the contribution from the electric current generated by the electron dynamics.
The reason why we could not have included such an important term is that we did not have a practical way to compute an oscillatory integral over the infinite interval involved in it. 
We shall report in this paper that there is an efficient method for such an integration \cite{Ooura1999,Ooura2005} based on so-called double exponential (DE) formula \cite{Takahashi1974}.


This paper is organized as follows. 
In Sec.~\ref{sec:te_qf}, we briefly review the equations of motion of QED field operators, and the canonical commutation or anticommutation relations among them.
In Sec.~\ref{sec:te_ca}, we derive the evolution equation for the electron creation and annihilation operators, and describe how the integrations for the retarded potential arise. 
In Sec.~\ref{sec:KjjKjE}, we show the results for the numerical integration of retarded potential terms using the DE formula.
Finally, Sec.~\ref{sec:conclusion} is devoted to our conclusion. 

%
\section{Time evolution equations for the quantum fields}  \label{sec:te_qf}
%

In this paper, there appear two types of quantum field operators, 
the four-component Dirac field operator $\hat{\psi}(x)$ for the electron and positron,  
and the $U(1)$ gauge field operator $\hat{A}_\mu(x)$ for the photon. 
(Incidentally, the approach to treat the electron as the two-component Schr\"{o}dinger field \cite{Tachibana2013}
is also developed in our group. See Refs.~\cite{Senami2013a,Senami2013b,Senami2014} for details.)
Their time evolutions are given by the Dirac equation and Maxwell equation.
Since we assume the canonical quantization formalism, the quantum operators satisfy the 
equal-time commutation or anticommutation relations.
Although these are textbook matters ({\it e.g.} \cite{WeinbergI}), we describe them briefly in this section to set up our notations. 
We work in the Gaussian systems of electromagnetic units. 
As for the physical constants, we use $c$ for the speed of light in vacuum, 
$\hbar$ for the reduced Planck constant, $e$ for the electron charge magnitude ($e > 0$),
and $m_e$ for the electron mass. 
The relativistic notations are as follows. 
The spacetime coordinate is expressed as $x = (x^\mu) = (x^0, x^i) = (ct, \vec{r})$, where the 
Greek index runs from 0 to 3 and the Latin index from 1 to 3. 
We adopt a convention that the repeated indices are summed, unless otherwise indicated. 
We use the metric tensor defined by $\eta_{\mu\nu} = {\rm diag}(1,-1,-1,-1) = \eta^{\mu\nu}$
for the transformation between contravariant and covariant vectors. 
The spacetime derivative is defined by $\partial_\mu = \frac{\partial}{\partial x^\mu} =  \left( \frac{1}{c}\frac{\partial}{\partial t}, \vec{\nabla} \right)$.
The gamma matrices are denoted by $\gamma^\mu$.

The Dirac equation in the covariant form is 
$i\hbar \gamma^\mu \hat{D}_{e \mu}(x) \hat{\psi}(x) = m_e c \hat{\psi}(x)$, where the gauge covariant derivative is 
defined by 
$\hat{D}_{e\mu}(x) = \partial_\mu + i\frac{Z_e e}{\hbar c} \hat{A}_\mu(x)$ with $Z_e = -1$, and this can be written as
\begin{eqnarray}
i\hbar \frac{\partial}{\partial t} \hat{\psi}(x) 
&=& \left\{-i\hbar  c  \gamma^0  \vec{\gamma} \cdot \vec{\nabla}
-(Z_e e) \gamma^0   \vec{\gamma} \cdot \hat{\vec{A}}(x)  
  + m_e c^2 \gamma^0  +  (Z_e e) \hat{A}_0(x) \right\} \hat{\psi}(x).  \label{eq:Dirac}
\end{eqnarray}
As we adopt the Coulomb gauge, $\vec{\nabla}\cdot \hat{\vec{A}}(x) = 0$, the Maxwell equation is given by
\begin{eqnarray}
\left( \frac{1}{c^2} \frac{\partial^2}{\partial t^2} - \nabla^2 \right) \hat{\vec{A}}(x)
=
\frac{4\pi}{c} \hat{\vec{j}}_T(x), \label{eq:Maxwell}
\end{eqnarray}
where $\hat{\vec{j}}_T(x)$ is the transversal part of the charge current density operator $\hat{\vec{j}}(x)$:
\begin{eqnarray}
\hat{\vec{j}}_T(x) = \hat{\vec{j}}(x)- \frac{1}{4\pi}  \vec{\nabla} \frac{\partial}{\partial t} \hat{A}_0(x). 
\label{eq:jT}
\end{eqnarray}
The scalar potential is given by 
\begin{eqnarray}
\hat{A}_0(ct, \vec{r}) = \int d^3\vec{s}\, \frac{\hat{\rho}(ct, \vec{s})}{|\vec{r}-\vec{s}|},
\label{eq:A_0}
\end{eqnarray}
where $\hat{\rho}(x)$ is the charge density operator, as the solution of the Poisson equation.
Here, $\hat{\rho}(x)$ and $\hat{\vec{j}}(x)$ are given in terms of the Dirac field operator, respectively, by
\begin{eqnarray}
\hat{\rho}(x)  &=& Z_e e \hat{\psi}^\dagger(x)\hat{\psi}(x), \label{eq:rho} \\
\hat{\vec{j}}(x) &=& Z_e\, e\, c\, \hat{\psi}^\dagger(x) \gamma^0 \vec{\gamma} \hat{\psi}(x), \label{eq:j}
\end{eqnarray}
where the dagger is used to express Hermite conjugate. 

The equal-time anticommutation relations for $\hat{\psi}(x)$ are
\begin{eqnarray}
\left\{  \hat{\psi}_\alpha(ct, \vec{r}), \hat{\psi}_\beta^\dagger (ct, \vec{s}) \right\} &=& \delta^{(3)}(\vec{r}-\vec{s}) \delta_{\alpha \beta},  \label{eq:Dirac_ac1} \\
\left\{  \hat{\psi}_\alpha(ct, \vec{r}), \hat{\psi}_\beta (ct, \vec{s}) \right\} &=& 0, \label{eq:Dirac_ac2} \\
\left\{  \hat{\psi}_\alpha^\dagger(ct, \vec{r}), \hat{\psi}_\beta^\dagger (ct, \vec{s}) \right\} &=& 0, \label{eq:Dirac_ac3}
\end{eqnarray}
where $\alpha, \beta = 1,...,4$ are spinor indices and the curly brackets are the anticommutator so that $\{ A, B \} = AB + BA$.
The equal-time commutation relations for  $\hat{\vec{A}}(x)$ consistent with the Coulomb gauge condition are
\begin{eqnarray}
\left[ \hat{A}^i(ct, \vec{r}), \hat{A}^j (ct, \vec{s}) \right] &=& 0,  \label{eq:EM_c1} \\
\left[ \hat{E}_T^i(ct, \vec{r}), \hat{E}_T^j (ct, \vec{s}) \right] &=& 0, \label{eq:EM_c2} \\
\frac{1}{4\pi c} \left[ \hat{A}^i(ct, \vec{r}), \hat{E}_T^j  (ct, \vec{s}) \right] 
&=& i\hbar \eta^{ij} \delta^{(3)}(\vec{r}-\vec{s})  + i\hbar \frac{\partial}{\partial r^i}\frac{\partial}{\partial r^j} \left( -\frac{1}{4\pi} \cdot \frac{1}{|\vec{r}-\vec{s}|} \right), \label{eq:EM_c3}
\end{eqnarray}
where 
the square brackets are the commutator so that $[A, B] = AB - BA$, and 
$\hat{\vec{E}}_T(x)$ is the transversal part of the electric field operator:
\begin{eqnarray}
\hat{\vec{E}}_T(x) = -\frac{1}{c} \frac{\partial \hat{\vec{A}}(x)}{\partial t}.
\end{eqnarray}
Finally, $\hat{\psi}(x)$ commutes with $\hat{\vec{A}}(x)$ at equal times:
\begin{eqnarray}
\left[  \hat{\psi}_\alpha(ct, \vec{r}), \hat{A}^i (ct, \vec{s}) \right] &=& 0.  \label{eq:D_EM_c}
\end{eqnarray}

%
\section{Time evolution of creation and annihilation operators}  \label{sec:te_ca}
%

Time evolution equations of the quantum fields given by Eqs.~\eqref{eq:Dirac} and \eqref{eq:Maxwell}
are very difficult to solve because not only they are nonlinear partial integro-differential equations
but also they are equations for non-commutative operators which obey the commutation and anticommutation relations Eqs.~\eqref{eq:Dirac_ac1}-\eqref{eq:EM_c3} and \eqref{eq:D_EM_c}.
In this section, we describe our prescription to make the equations more tractable.
Although there are some overlaps with the contents in Refs.~\cite{Ichikawa2013,Ichikawa2014},
we reproduce them in the reorganized form for the convenience of the readers.

Regarding $\hat{\psi}(x)$, we introduce creation and annihilation operators which carry the time dependence as follows \cite{Ichikawa2013,Ichikawa2014}.
The field operator is expanded by a set of four-component orthonormal functions $\psi_{n^a}(\vec{r})$ as
\begin{eqnarray}
\hat{\psi}(x) = \sum^{N_D}_{n=1} \sum_{a=\pm} \psi_{n^a}(\vec{r}) \hat{e}_{n^a}(t), \label{eq:psi_expand}
\end{eqnarray}
where $\int d^3{\vec{r}}\, \psi^\dagger_{n^a}(\vec{r}) \psi_{m^b}(\vec{r}) = \delta_{nm}\delta_{ab}$.
In our notation, $a=+$ and $a=-$ represent electron and positron respectively,
so that $\hat{e}_{n^+}$ is the electron annihilation operator and $\hat{e}_{n^-}$ is the positron creation operator. 
If the expansion functions are complete, the anticommutation relations Eqs.~\eqref{eq:Dirac_ac1}-\eqref{eq:Dirac_ac3}
lead to 
$\left\{ \hat{e}_{n^a}(t), \hat{e}^\dagger_{m^b}(t)   \right\}   = \delta_{nm} \delta_{ab}$,
$\left\{ \hat{e}_{n^a}(t), \hat{e}_{m^b}(t)   \right\}  = 0 $,
$\left\{ \hat{e}^\dagger_{n^a}(t), \hat{e}^\dagger_{m^b}(t)   \right\}  = 0$,
respectively.
Note that $N_D$, the number of the electron expansion function, has to be infinite to make the expansion function set complete.
As we can only use a finite set in numerical computation, thus obtained results should be interpreted as phenomena within the finite subspace.

As for $\hat{\vec{A}}(x)$, we first consider the integrated form of the Maxwell equation \eqref{eq:Maxwell} using the retarded Green function \cite{Tachibana2003,Tachibana2010}. 
The solution can be expressed by a sum of the radiation vector potential and the retarded potential as $\hat{\vec{A}}(ct, \vec{r}) = \hat{\vec{A}}_{\rm rad}(ct, \vec{r}) + \hat{\vec{A}}_A(ct, \vec{r})$, where
 \begin{eqnarray}
\hat{\vec{A}}_{\rm rad}(ct, \vec{r})
 &=& \frac{\sqrt{4\pi \hbar^2 c}}{\sqrt{(2\pi\hbar)^3}} \sum_{\sigma=\pm1} \int \frac{d^3 \vec{p}}{\sqrt{2p^0}}
\bigg[ \hat{a}_{\vec{p}_\sigma} \vec{e}(\vec{p},\sigma) e^{-i c p^0 t/\hbar} e^{i \vec{p}\cdot \vec{r} /\hbar} 
+ \hat{a}^\dagger_{\vec{p}_\sigma} \vec{e}^{\, *}(\vec{p},\sigma) e^{i c p^0 t/\hbar} e^{-i \vec{p}\cdot \vec{r} /\hbar} \bigg] , \label{eq:A_rad} \nonumber \\
\\
\hat{\vec{A}}_A(ct, \vec{r}) 
&=&  \frac{1}{c} \int d^3\vec{s}\, \frac{\hat{\vec{j}}_T(c u, \vec{s})}{|\vec{r}-\vec{s}|},  \quad 
u = t - \frac{|\vec{r}-\vec{s}|}{c}.   \label{eq:A_A}
\end{eqnarray}
In Eq.~\eqref{eq:A_rad}, $\vec{p}$ and $\sigma$ denote the photon momentum and helicity respectively.
The usual dispersion relation $p^0 = |\vec{p}|$ holds and the polarization vector $\vec{e}(\vec{p},\sigma)$ satisfies 
$\vec{p} \cdot \vec{e}(\vec{p},\sigma)=0$, 
$\sum_{\sigma=\pm 1} e^i(\vec{p},\sigma) e^{*j}(\vec{p},\sigma) = -\eta^{ij} - \frac{p^i p^j}{|\vec{p}|^2}$,
and $\sum_{k=1}^3 e^k(\vec{p},\sigma) e^{*k}(\vec{p},\tau) = \delta_{\sigma \tau}$.
The photon annihilation operator $\hat{a}_{\vec{p}_\sigma}$ 
satisfies the commutation relations $[\hat{a}_{\vec{p}_\sigma}, \hat{a}_{\vec{q}_\tau}] = [\hat{a}^\dagger_{\vec{p}_\sigma}, \hat{a}^\dagger_{\vec{q}_\tau}] = 0$ and $[\hat{a}_{\vec{p}_\sigma}, \hat{a}^\dagger_{\vec{q}_\tau}] = \delta^{(3)}(\vec{p}-\vec{q}) \delta_{\sigma \tau}$ to be consistent with Eqs.~\eqref{eq:EM_c1}-\eqref{eq:EM_c3}.
Note that $\hat{a}_{\vec{p}_\sigma}$ is time-independent. 

The integration of Eq.~\eqref{eq:A_A} contains the retarded time $u$, which reflects the fact that the speed of light (the maximum speed at which the information can be transmitted) is finite and we only use the information from the past.  
Since $u$ depends on space variables, it is difficult to perform the integration in this form.
We rewrite it using the delta function formulae
with the causality ($\hat{\vec{j}}_T(cu, \vec{r})=0$ for $u>t$) and the initial condition ($\hat{\vec{j}}_T(cu, \vec{r})=0$ for $u<t_0$) as 
\begin{eqnarray}
\hat{\vec{A}}_A(ct, \vec{r}) &=& 
\frac{1}{c^2 \pi} \int_{t_0}^{t }du'  \int_{-\infty}^{\infty}d\alpha \int d^3\vec{s}\, \hat{\vec{j}}_T(c u', \vec{s}) \exp \left[ i\alpha \left\{ (t-u')^2 - \frac{(\vec{r}-\vec{s})^2}{c^2} \right\} \right], 
\label{eq:A_A_2}
\end{eqnarray}
separating the time and space variables \cite{Tachibana2013, Tachibana2015}.
The integration with respect to $u'$ represents the accumulation of contributions from past data, and 
the integration with respect to $\alpha$ sweeps out the non-causal data.
 As for the initial condition, we note that we set the Cauchy problem of QED by assuming the synchronization of the clocks at different space points at $t = t_0$,
 when canonical quantization is performed with the definition of the vacuum ket vector $| 0 \rangle$. 
Hence, the vacuum and field operators are not defined for $t < t_0$. 

We can express Eq.~\eqref{eq:A_A_2} with the creation and annihilation operators by using Eqs.~\eqref{eq:jT}, \eqref{eq:A_0}, \eqref{eq:rho}, \eqref{eq:j}, and \eqref{eq:psi_expand} as
\begin{eqnarray}
\hat{A}^k_A(ct, \vec{r})
&=& 
\frac{1}{c^2 \pi}  \sum_{p,q=1}^{N_D} \sum_{c,d=\pm}  \int_{t_0}^{t }du'  
\left\{ 
K^k_{j, p^c q^d}(\vec{r}; t-u') \hat{\cal E}_{p^c q^d}(u')
+  K^k_{E, p^c q^d}(\vec{r}; t-u')   \frac{d \hat{\cal E}_{p^c q^d}}{dt}(u') 
\right\},  \label{eq:A_A_3} \nonumber \\
\end{eqnarray}
where we define the frequently encountered combination of the operator $\hat{\cal E}_{p^c q^d}(t)= \hat{e}^\dagger_{p^c}(t) \hat{e}_{q^d}(t)$, and $c$-number integrals
\begin{eqnarray}
K^k_{j, p^c q^d}(\vec{r}; t-u') &=& \int_{-\infty}^{\infty}d\alpha\,  I^k_{j, p^c q^d}(\vec{r}; \alpha) \exp \left( i\alpha (t-u')^2 \right),   \\
K^k_{E, p^c q^d}(\vec{r}; t-u') &=& \int_{-\infty}^{\infty}d\alpha\,  I^k_{E, p^c q^d}(\vec{r}; \alpha) \exp \left( i\alpha (t-u')^2 \right). 
\end{eqnarray}
Here, the integrals  $I^k_{j, p^c q^d}(\vec{r}; \alpha)$ and $I^k_{E, p^c q^d}(\vec{r}; \alpha)$ are defined as
\begin{eqnarray}
I^k_{j, p^c q^d}(\vec{r}; \alpha)
&=&   \int d^3\vec{s}\, j^k_{p^c q^d}(\vec{s}) \exp \left( -i\alpha \frac{(\vec{r}-\vec{s})^2}{c^2} \right),  \label{eq:Ij}  \\
 I^k_{E, p^c q^d}(\vec{r}; \alpha) 
&=&  \int d^3\vec{s}\, E^k_{p^c q^d}(\vec{s})  \exp \left( -i\alpha \frac{(\vec{r}-\vec{s})^2}{c^2} \right), \label{eq:IE} 
\end{eqnarray}
respectively using following functions
\begin{eqnarray}
j^k_{p^c q^d}(\vec{s}) 
&=& 
Z_e\, e\,c \left[ \psi_{p^c}^\dagger(\vec{s}) \gamma^0 \gamma^k \psi_{q^d}(\vec{s})  \right],  \\
E^{k}_{p^c q^d}(\vec{s}) 
&=& 
-\frac{Z_e e}{4\pi} \int d^3\vec{t}\, \psi_{p^c}^\dagger(\vec{t}) \psi_{q^d}(\vec{t})  \frac{(\vec{t}-\vec{s})^k}{|\vec{t}-\vec{s}|^3}.   \label{eq:ef_int} 
\end{eqnarray}

Finally, we obtain the time evolution equation for the creation and annihilation operators by
substituting the vector potential, \eqref{eq:A_rad}  and \eqref{eq:A_A_3}, into the Dirac equation \eqref{eq:Dirac}, and applying the expansion \eqref{eq:psi_expand} as
\begin{eqnarray}
i\hbar \frac{d \hat{e}_{n^a}}{d t}(t) &=& 
  \sum_{m=1}^{N_D}  \sum_{b=\pm}  (T_{n^a m^b} + M_{n^a m^b} )\hat{e}_{m^b}(t) 
  +    \sum_{m, p, q=1}^{N_D} \sum_{b, c, d=\pm}(n^a m^b | p^c q^d) \hat{\cal E}_{p^c q^d}(t)   \hat{e}_{m^b}(t)  
  \nonumber \\
&-&\frac{1}{c^3 \pi} \sum_{m,p,q=1}^{N_D}  \sum_{b,c,d=\pm}  \int_{t_0}^{t }du' 
\Bigg\{ K_{jj, n^a m^b p^c q^d}(t-u') \hat{\cal E}_{p^c q^d}(u')  \nonumber \\
& & +K_{jE, n^a m^b p^c q^d}(t-u')  \frac{d \hat{\cal E}_{p^c q^d}}{dt}(u') 
\Bigg\} \hat{e}_{m^b}(t)  
   \nonumber \\
&-& \sqrt{\frac{1}{2\pi^2 \hbar c}}
 \sum_{m=1}^{N_D}  \sum_{b=\pm}  \sum_{\sigma=\pm1} \int \frac{d^3 \vec{p}}{\sqrt{2p^0}} 
\left[ 
 {\cal F}_{n^a m^b \vec{p}_\sigma}(t) \hat{a}_{\vec{p}_\sigma}   + {\cal F}^*_{m^b n^a \vec{p}_\sigma}(t)   \hat{a}^\dagger_{\vec{p}_\sigma}  
\right]  \hat{e}_{m^b}(t) , \label{eq:dedt} \nonumber \\
\end{eqnarray}
where we define the kinetic energy integral
$T_{n^a m^b} = -i\hbar  c   \int d^3\vec{r}\, \psi_{n^a}^\dagger(\vec{r}) \gamma^0 \gamma^k   \partial_k \psi_{m^b}(\vec{r})$,
the mass energy integral
$M_{n^a m^b} = m_e c^2 \int d^3\vec{r}\, \psi_{n^a}^\dagger(\vec{r})  \gamma^0  \psi_{m^b}(\vec{r})$,
the two-electron-repulsion integral
$(n^a m^b | p^c q^d) =
(Z_e e)^2 \int d^3\vec{r}\, d^3\vec{s}\, \psi_{n^a}^\dagger(\vec{r}) \psi_{m^b}(\vec{r}) 
\frac{1}{|\vec{r}-\vec{s}|} \psi_{p^c}^\dagger(\vec{s}) \psi_{q^d}(\vec{s})$,
and
${\cal F}_{n^a m^b \vec{p}_\sigma}(t) = \sum_{k=1}^3  e^k(\vec{p},\sigma) e^{-i c p^0 t/\hbar} F^k_{n^a m^b}(\vec{p})$,
which is defined using the Fourier transform of $j^k_{n^a m^b}(\vec{r})$,
$F^k_{n^a m^b}(\vec{p})  \equiv  \int d^3\vec{r}\, j^k_{n^a m^b}(\vec{r})  e^{i \vec{p}\cdot \vec{r} /\hbar}$.
In Eq.~\eqref{eq:dedt}, we also define integrals which originate from the interaction between the electronic current and retarded potential as
\begin{eqnarray}
K_{jj, n^a m^b p^c q^d}(t-u') &=& \int_{-\infty}^{\infty}d\alpha\, I_{jj, n^a m^b p^c q^d}(\alpha) \exp \left( i\alpha (t-u')^2 \right), \label{eq:Kjj} \\
K_{jE, n^a m^b p^c q^d}(t-u') &=& \int_{-\infty}^{\infty}d\alpha\, I_{jE, n^a m^b p^c q^d}(\alpha) \exp \left( i\alpha (t-u')^2 \right),  \label{eq:KjE} 
\end{eqnarray}
where
\begin{eqnarray}
I_{jj, n^a m^b p^c q^d}(\alpha) &=& \sum_{k=1}^3 \int d^3\vec{r}\, j^k_{n^a m^b}(\vec{r}) I^k_{j, p^c q^d}(\vec{r}; \alpha) \\
&=& \sum_{k=1}^3 \int d^3\vec{r}\, d^3\vec{s}\,  j^k_{n^a m^b}(\vec{r}) j^k_{p^c q^d}(\vec{s}) \exp \left( -i\alpha \frac{(\vec{r}-\vec{s})^2}{c^2} \right),  \label{eq:Ijj} \\
 I_{jE, n^a m^b p^c q^d}(\alpha) &=&  \sum_{k=1}^3 \int d^3\vec{r}\, j^k_{n^a m^b}(\vec{r}) I^k_{E, p^c q^d}(\vec{r}; \alpha) \\
 &=& \sum_{k=1}^3 \int d^3\vec{r}\, d^3\vec{s}\,  j^k_{n^a m^b}(\vec{r}) E^k_{p^c q^d}(\vec{s})  \exp \left( -i\alpha \frac{(\vec{r}-\vec{s})^2}{c^2} \right). \label{eq:IjE}
\end{eqnarray}


%
\section{Numerical integration of retarded potential terms}  \label{sec:KjjKjE}
%
As is described in the previous section, we have rewritten the coupled Maxwell-Dirac field equations
into the evolution equation for the electron creation and annihilation operators, Eq.~\eqref{eq:dedt}.
That is, we have succeeded in converting the partial integro-differential equations for the quantum operators, which depend on space-time coordinate,
into the ordinary integro-differential equation for the creation and annihilation operators, which carry only the time variable.
While some of the coefficients of the evolution equation, $T_{n^a m^b}$, $M_{n^a m^b}$, $(n^a m^b | p^c q^d)$, and $F^k_{n^a m^b}(\vec{p})$, are well-known molecular integrals, $K_{jj, n^a m^b p^c q^d}$ and $K_{jE, n^a m^b p^c q^d}$ do not appear in the quantum chemistry computation using the electrostatic Hamiltonian. Below, we describe how we may practically compute them.
Our computation assumes that the expansion functions $\psi_{n^a}(\vec{r})$ for the Dirac field operator to be expressed by a linear combination of Gaussian type orbitals. We use the DIRAC program package \cite{DIRAC12} to generate $\psi_{n^a}(\vec{r})$.  
The results are reported in the atomic units, in which the speed of light is $c= 137.035999679$ and 1\,a.u. of time equals to $24.19$\,as.

We begin with the computation of $K_{jj, n^a m^b p^c q^d}$ defined by Eq.~\eqref{eq:Kjj}. As this integral depends on the parameter $t-u'$, which represents the difference between the present time $t$ and the past time $u'$, if we want to perform a simulation from $t=t_0$ to $t=t_{end}$, we need $K_{jj, n^a m^b p^c q^d}$ with $t-u'$ ranging from $0$ to $t_{end}$. 
Therefore, unless $t-u' =0$, $K_{jj, n^a m^b p^c q^d}$ is an oscillatory integral over the infinite interval, 
which is in general difficult to make converge. 
In fact, although we have tried to use the Romberg integration of improper integrals such as found in Ref.~\cite{NumericalRecipe} and the fast Fourier transform, these methods turn out to be not practical. 
Then, we have noticed that an efficient method for such a Fourier-type integral (the value of a Fourier transform at a particular point) has been developed in Ref.~\cite{Ooura1999,Ooura2005} based on the 
DE formula \cite{Takahashi1974}, and 
its implementation is made publicly available by the developer \cite{OouraMSP}.
We perform the integration with respect to $\alpha$ in Eq.~\eqref{eq:Kjj} by using the Ooura's code, and the six-dimesional integration with respect to spatial coordinates in Eq.~\eqref{eq:Ijj} by using the analytic formula explained in the Appendix \ref{sec:molint}. 

\begin{figure}
\begin{center}
\includegraphics[width=12cm]{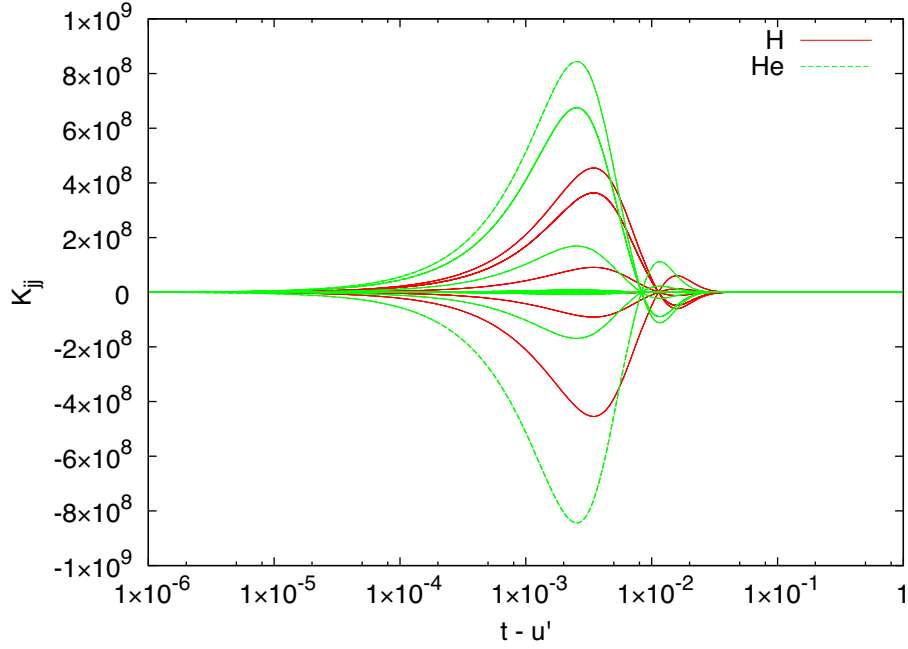}
\caption{The real part of $K_{jj, n^a m^b p^c q^d}$ as a function of $t-u'$. All the components are plotted.
}
\label{fig:Kjj}
\end{center}
\end{figure}

The numerical results for $K_{jj, n^a m^b p^c q^d}$ as a function of $t-u'$ are shown in Fig.~\ref{fig:Kjj}.
The expansion functions are generated by solving the  four-component Dirac equation with the Dirac-Coulomb Hamiltonian by the Hartree-Fock method and STO-3G basis set. 
We show the results for two types of expansion functions, which are respectively generated using H and He atoms.
We note that, for these atoms in this basis set, there are two orbitals ($N_D=2$) for electron and positron respectively taking into account the Kramers partners.
In Fig.~\ref{fig:Kjj}, all the ($4^4=256$) components of $K_{jj, n^a m^b p^c q^d}$ are plotted so that multiple lines are shown for each H and He. 
Since the imaginary part of $K_{jj, n^a m^b p^c q^d}$ is found to be zero for all the components, the real part is plotted. 
In the limiting case of $t-u' =0$, we can analytically evaluate the integral to be zero (see Appendix \ref{sec:anaKjj}).
Also, for $t-u' \rightarrow \infty$, $K_{jj, n^a m^b p^c q^d}$ becomes zero owing to the Riemann-Lebesgue lemma. 
We see that our numerical results reproduce these behaviors at the limiting cases.  
We also notice that the value of $K_{jj, n^a m^b p^c q^d}$ decreases rapidly when $t-u'$ is larger than around $10^{-2}$.
This can be attributed to the fact that our expansion functions extend over about 1\,{a.u.}~and so does the source of the retarded potential. 
Note that $K_{jj, n^a m^b p^c q^d}(t-u')$ expresses the contribution to the retarded potential at the past time $u'$, $t-u'$ before the present time $t$.
Since the information is transmitted at the speed of light, there should be no contribution from the time before approximately $1 / c$, that is, $(t-u') > O(10^{-2})$.
This is supported by the result that $K_{jj, n^a m^b p^c q^d}(t-u')$ for He has a peak at smaller $t-u'$ than that of H,
which is consistent with the less extended orbital of He than H (the exponent of the He basis set is about 1.9 times larger than that of H). 

\begin{figure}
\begin{center}
\includegraphics[width=12cm]{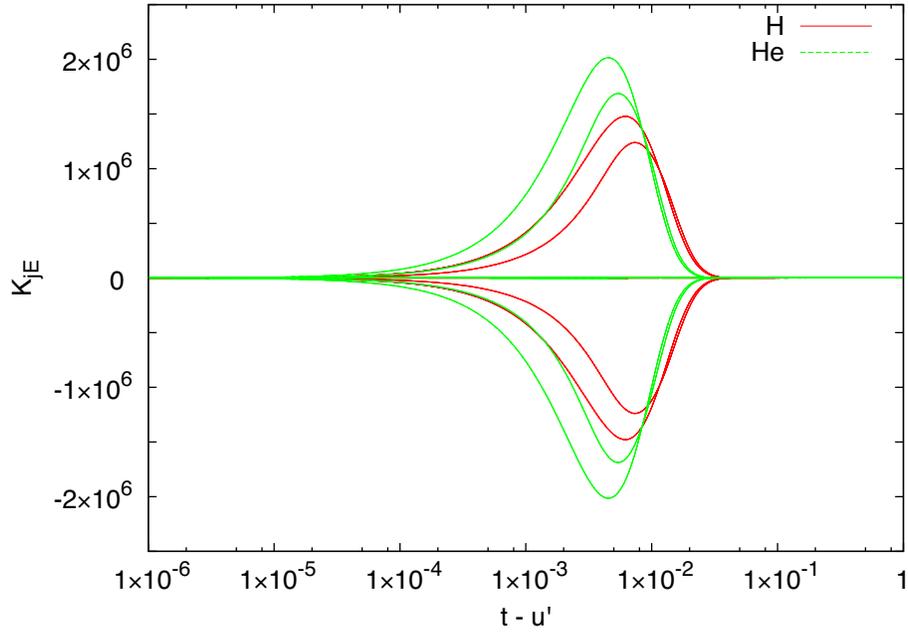}
\caption{The imaginary part of $K_{jE, n^a m^b p^c q^d}$ as a function of $t-u'$. All the components are plotted.
}
\label{fig:KjE}
\end{center}
\end{figure}

The computation of $K_{jE, n^a m^b p^c q^d}$ defined by Eq.~\eqref{eq:KjE} is performed using the same technique as described above, and the result is shown in Fig.~\ref{fig:KjE}.
Since the real part of $K_{jE, n^a m^b p^c q^d}$ is found to be zero for all the components, the imaginary part is plotted. 
Fig.~\ref{fig:KjE} exhibits a similar pattern to Fig.~\ref{fig:Kjj}, and this can be interpreted in a similar manner to that of $K_{jj, n^a m^b p^c q^d}$ as described above. 
We, however, have to remember that the integrand of $K_{jE, n^a m^b p^c q^d}$ contains a factor $I_{jE, n^a m^b p^c q^d}(\alpha)$ which diverges at $\alpha=0$ like a delta function (Eq.~\eqref{eq:IjE0} in Appendix \ref{sec:molint}).
This contributes to $K_{jE, n^a m^b p^c q^d}$ as an indefinite constant term which in general depends on the component, but not on the spacetime coordinate.  
The result shown in Fig.~\ref{fig:KjE} omits this possible contribution. 
We may determine this constant term by looking at other quantity such as the Hamiltonian operator but 
it is beyond the scope of the current paper. 

%
\section{Conclusion}  \label{sec:conclusion}
%

In this paper, we have discussed the method to compute the integrals, denoted by $K_{jj}$ and $K_{jE}$, which appear in the retarded potential term for a real-time simulation based on QED.
We have shown that the oscillatory integrals over the infinite interval involved in them can be efficiently performed by the method developed by Ooura and Mori based on the DE formula. 
Now, we can set almost all the coefficients for the evolution equation of the electron creation and annihilation operators, Eq.~\eqref{eq:dedt}.
We, however, also have found that there seems to be an indefinite constant contribution to $K_{jE}$, which stems from the delta-function-like singularity in its integrand. 
How we may set the constant is not known at this stage, and we shall look for a way by 
investigating other quantity such as the Hamiltonian operator or using more sophisticated mathematical technique to treat the singularity. 

Even if we find a way to fix the constant and obtain $K_{jE}$, there will be several issues in solving Eq.~\eqref{eq:dedt}.
One of them is a reasonable matrix representation of the creation and annihilation operators.
This may be determined by using a method of constructing the basis ket vectors using newly found $b$-photon, $f$-electron, and $f^c$-positron field operators \cite{Tachibana2015}.
As they work for interacting theory, we do not have to invoke asymptotic fields to define a Fock space on which the creation and annihilation operators act, and non-perturbative formulation is possible. 
Our next task would be to connect $\hat{e}_{n^a}(t)$ in this paper and $b$-photon, $f$-electron, and $f^c$-positron field operators introduced in Ref.~\cite{Tachibana2015}. 

Further issue is that, in addition to the time evolution of the quantum operators, we have to solve the other side of the dual Cauchy problem of QED as mentioned in the introduction.
That is, the time evolution of the wavefunctions. 
In quantum field theory, we have to deal with the infinite series of wavefunctions, each of them representing a certain combination of fixed numbers of electrons, positrons and photons \cite{Tachibana2015}.
This is in contrast to the quantum mechanics of point particles, in which the particle number is conserved and only one wavefunction is needed. 
Such huge increase in the degree of freedom in quantum field theoretic computation would require some reduction techniques for a practical numerical implementation. 

Our quantum field theoretic formulation for the real-time simulation based on QED requires a lot of ingredients which differ from those in the quantum mechanics of point particles based on the electrostatic Hamiltonian.  
In this paper, we have taken one step further to achieve our goal by showing the practical computational method for the integrations in the retarded potential in QED.

\noindent 
\section*{Acknowledgment}
Theoretical calculations were partly performed using Research Center for Computational Science, Okazaki, Japan.
This work was supported by JSPS KAKENHI Grant Number 25410012 and 26810004.
M.~F. is supported by the Grant-in-Aid for JSPS Fellows (Grant Number 14J02866).



\appendix

%
\section{Molecular integral formulae for retarded potential term}  \label{sec:molint}
%
In this appendix, we summarize the gaussian integral formulae to compute
retarded potential terms.
Specifically, we describe formulae to compute 
$I_{jj, n^a m^b p^c q^d}(\alpha)$, and  $I_{jE, n^a m^b p^c q^d}(\alpha)$, which are respectively defined by 
Eqs.~\eqref{eq:Ijj}, and \eqref{eq:IjE}.
Our formulae and derivation here are based on a method described in Ref.~\cite{McMurchie1978}, and we mostly follow their notations.
Although there are significant overlaps in this section with the appendix of Ref.~\cite{Ichikawa2014},
we reproduce them with the typos fixed for the convenience of the readers.

Since $I_{jj, n^a m^b p^c q^d}(\alpha)$ and $I_{jE, n^a m^b p^c q^d}(\alpha)$ are four-center integrals, we need to compute  basic two-electron integrals 
\begin{eqnarray}
& &\left[ NLM | \theta | N' L' M' \right] \nonumber \\
&& = \left(\frac{\partial}{\partial P_x}\right)^N \left(\frac{\partial}{\partial P_y}\right)^L \left(\frac{\partial}{\partial P_z}\right)^M
\left(\frac{\partial}{\partial Q_x}\right)^{N'} \left(\frac{\partial}{\partial Q_y}\right)^{L'} \left(\frac{\partial}{\partial Q_z}\right)^{M'}\left[ 000 | \theta | 000 \right],
\end{eqnarray}
where
\begin{eqnarray}
[000|\theta|000] = \int d^3\vec{r}\,  d^3\vec{s}\,  \exp\left(-\alpha_P  |\vec{r} - \vec{P} |^2  \right) \exp\left(-\alpha_Q  |\vec{s} - \vec{Q} |^2  \right) \theta(\vec{r}, \vec{s}),  \label{eq:000theta000}
\end{eqnarray}
with $\theta(\vec{r}, \vec{s})$ being
\begin{eqnarray}
\theta_{jj}(\vec{r}, \vec{s}; \alpha) \equiv \exp \left( -i\alpha  \frac{|\vec{r}-\vec{s}|^2}{c^2} \right),  \label{eq:thetajj}
\end{eqnarray}
for $I_{jj, n^a m^b p^c q^d}(\alpha)$, and 
\begin{eqnarray}
\theta^k_{jE}(\vec{r}, \vec{s}; \alpha) \equiv \int d^3\vec{t}\,   
\frac{(\vec{s}-\vec{t})^k}{|\vec{s}-\vec{t}|^3} \exp \left( -i\alpha  \frac{|\vec{r}-\vec{t}|^2}{c^2} \right), \label{eq:thetajE}
\end{eqnarray}
for $I_{jE, n^a m^b p^c q^d}(\alpha)$. 

In the case of  $\alpha \neq 0$, it is shown that
\begin{eqnarray}
\left[ 000 | \theta_{jj} | 000 \right] &=& \pi^3 B^{-3/2} \exp \left( -\alpha_T |\vec{D}|^2 \right), 
\label{eq:000thetajj000} \\
\left[ 000 | \theta^k_{jE} | 000 \right] &=& -4 \pi^4 B^{-3/2} F_1(\alpha_T  |\vec{D}|^2) D^k,
\label{eq:000thetajE000}
\end{eqnarray}
where 
$\vec{D} = \vec{P} -\vec{Q}$,
$A = i\alpha/c^2$,
$B =  A(\alpha_P + \alpha_Q) +  \alpha_P \alpha_Q$,
$C = \alpha_P \alpha_Q  A$,
$\alpha_T = \left(  1/\alpha_P + 1/\alpha_Q + 1/A \right)^{-1} = C/B$,
and
\begin{eqnarray}
F_j(T) = \int_0^1 u^{2j} \exp\left( -Tu^2 \right) du.
\end{eqnarray}
%

It is  straightforward to differentiate Eqs.~\eqref{eq:000thetajj000} and \eqref{eq:000thetajE000} to derive $\left[ NLM | \theta | N' L' M' \right]$.
As for $\theta_{jj}$, 
\begin{eqnarray}
\left[ NLM | \theta_{jj} | N' L' M' \right] &=& 
\pi^3 B^{-3/2} \exp \left( -\alpha_T |\vec{D}|^2 \right) \alpha_T^{\frac{N+L+M+N'+L'+M'}{2}}(-1)^{N+L+M} \nonumber \\
&\times& H_{N+N'}(\alpha_T^{1/2} D_x) H_{L+L'}(\alpha_T^{1/2} D_y) H_{M+M'}(\alpha_T^{1/2} D_z),
\end{eqnarray}
where $H_n(x)$ is a Hermite polynomial of degree $n$.
As for $\theta_{jE}$, 
\begin{eqnarray}
\hspace{-2cm}
\left[ NLM | \theta^x_{jE} | N' L' M' \right] &=& 
  -4 \pi^4 B^{-3/2}  (-1)^{N'+L'+M'}  \nonumber \\ 
  &\times&  \left\{ D_x  \tilde{R}_{N+N', L+L', M+M'} + (N+N') \tilde{R}_{N+N'-1, L+L', M+M'} \right\}, \\
\left[ NLM | \theta^y_{jE} | N' L' M' \right] &=&  
  -4 \pi^4 B^{-3/2}  (-1)^{N'+L'+M'} \nonumber \\
  &\times & \left\{ D_y  \tilde{R}_{N+N', L+L', M+M'} + (L+L') \tilde{R}_{N+N', L+L'-1, M+M'} \right\}, \\
\left[ NLM | \theta^z_{jE} | N' L' M' \right] &=&  
  -4 \pi^4 B^{-3/2}  (-1)^{N'+L'+M'}\nonumber \\  
  &\times & \left\{ D_z  \tilde{R}_{N+N', L+L', M+M'} + (M+M') \tilde{R}_{N+N', L+L', M+M'-1} \right\},
\end{eqnarray}
where we have defined
\begin{eqnarray}
\tilde{R}_{NLM} 
=  \left(\frac{\partial}{\partial D_x}\right)^N \left(\frac{\partial}{\partial D_y}\right)^L \left(\frac{\partial}{\partial D_z}\right)^M F_1(T), 
\end{eqnarray}
with $T = \alpha_T (D_x^2+D_y^2+D_z^2)$.
For generating a table of all $\tilde{R}_{NLM}$ up to some maximum $N+L+M$, recursion relations discussed in Ref.~\cite{McMurchie1978} can be applied. 
In particular, we can use the recursion relation for the more general integral $R_{NLMj}$, 
\begin{eqnarray}
R_{NLMj} &=&
 (-\alpha_T^{1/2})^{N+L+M}(-2\alpha_T)^j \nonumber \\
&& \times \int_0^1 u^{N+L+M+2j} H_N(\alpha_T^{1/2} D_x u)H_L(\alpha_T^{1/2} D_y u)H_M(\alpha_T^{1/2} D_z u) e^{-Tu^2} du,
\end{eqnarray}
through the relation
$\tilde{R}_{NLM} = -R_{NLM 1}/(2\alpha_T)$.
The details of the recursion relations are found in Ref.~\cite{McMurchie1978}, and our code used in this paper follows their prescriptions.

Let us now consider the case of $\alpha = 0$.
The expressions we have derived above for the case of $\alpha \ne 0$, Eqs.~\eqref{eq:000thetajj000} and \eqref{eq:000thetajE000}, are finite at $\alpha=0$.
As $\alpha=0$ implies $\alpha_T = 0$, 
the right-hand-sides of Eqs.~\eqref{eq:000thetajj000} and \eqref{eq:000thetajE000} respectively become 
$\pi^3 (\alpha_P \alpha_Q)^{-3/2}$ and 
$-(4/3) \pi^4 (\alpha_P \alpha_Q)^{-3/2} D^k$.
We, however, have to set $\alpha = 0$ before the integration, that is, to use 
$\theta_{jj}(\vec{r}, \vec{s}; \alpha=0) = 1$ and 
$\theta^k_{jE}(\vec{r}, \vec{s}; \alpha=0) = \int d^3\vec{t}\,   
\frac{(\vec{s}-\vec{t})^k}{|\vec{s}-\vec{t}|^3}$.
As for $\theta_{jj}$, Eq.~\eqref{eq:000theta000} becomes the product of two overlap integrals, which leads to $(\pi/\alpha_P)^{3/2} \cdot (\pi/\alpha_Q)^{3/2}$, giving the same result as setting $\alpha=0$ in the right-hand-side expression of Eq.~\eqref{eq:000thetajj000}.
As for $\theta_{jE}$, Eq.~\eqref{eq:000theta000} turns out to be divergent as
\begin{eqnarray}
 -4 \pi^4     (\alpha_P \alpha_Q)^{-3/2} i  \times \lim_{\vec{l} \rightarrow \vec{0}} e^{-l^2/(4\alpha_Q)} e^{+ i \vec{l}\cdot \vec{Q}} \, \frac{l^k}{l^2}.
\end{eqnarray}
Therefore, $I_{jE, n^a m^b p^c q^d}(\alpha)$ has a delta-function-like structure as
\begin{eqnarray}
I_{jE, n^a m^b p^c q^d}(\alpha) = \bar{I}_{jE, n^a m^b p^c q^d}(\alpha)+ C_{n^a m^b p^c q^d}\, \delta(\alpha),
\label{eq:IjE0}
\end{eqnarray}
 where $\bar{I}_{jE, n^a m^b p^c q^d}(\alpha)$ is a part constructed using the expression of Eq.~\eqref{eq:000thetajE000} for whole range of $\alpha$ including $\alpha=0$, and $C_{n^a m^b p^c q^d}$ is an indefinite constant.

%
\section{Analytic computation of $K_{jj, n^a m^b p^c q^d}(t-u')$ for $t-u' = 0$}  \label{sec:anaKjj}
%
$K_{jj, n^a m^b p^c q^d}(t-u')$ is defined by Eq.~\eqref{eq:Kjj} and, when $t-u' = 0$, the integration can be done analytically as follows.
\begin{eqnarray}
 K_{jj, n^a m^b p^c q^d}(0)
 &=& \int_{-\infty}^{\infty}d\alpha\, I_{jj, n^a m^b p^c q^d}(\alpha),  \\
 &=& 
 \sum_{k=1}^3 \int_{-\infty}^{\infty}d\alpha\, \int d^3\vec{r}\, d^3\vec{s}\,  j^k_{n^a m^b}(\vec{r}) j^k_{p^c q^d}(\vec{s}) \exp \left( -i\alpha \frac{(\vec{r}-\vec{s})^2}{c^2} \right), \\
 &=&
 2\pi c^2 \sum_{k=1}^3  \int d^3\vec{r}\,  d^3\vec{s}\,  j^k_{n^a m^b}(\vec{r})  j^k_{p^c q^d}(\vec{s})  \delta\left( (\vec{r}-\vec{s})^2 \right), \\
&=&
 2\pi c^2 \sum_{k=1}^3  \int d^3\vec{r}\, \int_0^\infty dR \int_0^{\pi} d\theta  \int_0^{2\pi} d\phi R^2 \sin\theta  \,  j^k_{n^a m^b}(\vec{r})  j^k_{p^c q^d}(\vec{r} + \vec{R} )  \delta\left(R^2 \right), \nonumber \\
\end{eqnarray}
where we have defined $\vec{R} = \vec{s} - \vec{r}$ and  converted the integration over $\vec{s}$ into the integration over the polar coordinate $(R, \theta, \phi)$ centered at $\vec{r}$. Then, this expression becomes zero upon the integration over $R$. 
Similarly, we can show $K_{jE, n^a m^b p^c q^d}(0) = 0$.


\newpage

\end{document}